\documentclass[10pt, conference, compsocconf]{IEEEtran}
\usepackage{amsmath}

\ifCLASSINFOpdf
  \usepackage[pdftex]{graphicx}
\else
  \usepackage[dvips]{graphicx}
\fi
\usepackage{color}

\usepackage{algorithm}
\usepackage{algorithmicx}
\usepackage{algpseudocode}
\usepackage[tight,footnotesize]{subfigure}

\usepackage{stfloats}
\begin{document}
%
\title{Optimising energy and overhead for large parameter space simulations}


\author{\IEEEauthorblockN{Alexander J. M. Kell, Matthew Forshaw, A. Stephen McGough}
\IEEEauthorblockA{School of Computing\\
Newcastle University\\
Newcastle, UK\\
\{a.kell2, matthew.forshaw, stephen.mcgough\}@newcastle.ac.uk}
	}

\maketitle
\begin{abstract}
Many systems require optimisation over multiple objectives, where objectives are characteristics of the system such as energy consumed or increase in time to perform the work. Optimisation is performed by selecting the `best' set of input parameters to elicit the desired objectives. However, the parameter search space can often be far larger than can be searched in a reasonable time. Additionally, the objectives are often mutually exclusive -- leading to a decision being made as to which objective is more important or optimising over a combination of the objectives. This work is an application of a Genetic Algorithm to identify the Pareto frontier for finding the optimal parameter sets for all combinations of objectives. A Pareto frontier can be used to identify the sets of optimal parameters for which each is the `best' for a given combination of objectives -- thus allowing decisions to be made with full knowledge. We demonstrate this approach for the HTC-Sim simulation system in the case where a Reinforcement Learning scheduler is tuned for the two objectives of energy consumption and task overhead. Demonstrating that this approach can reduce the energy consumed by ${\sim}$36 \% over previously published work without significantly increasing the overhead.
\end{abstract}


%
\IEEEpeerreviewmaketitle

\section{Introduction}
There is a strong desire to model real-world systems through computer simulation -- a software system which replicates the salient features of the real-world system. This permits `what if' analysis, where one desires to know how the real-world system will be affected by changes in environment or policy. This is especially important when the proposed changes to the real system would be unpalatable to perform -- such as costing too much, having significant impact or potentially causing a degradation of service. 

In recent years the concept of the `digital twin', a simulation of a specific instance rather than a generic type of system, has emerged. Allowing `what if' analysis of the digital twin which can then be applied to the real system. For example, optimising the parameters controlling how the system performs. Traditionally this would be very difficult to perform on the real system due to fears that changes could have unforeseen detrimental impacts. However, by making the changes to the digital twin we remove this risk and can perform many simulations faster-than-real-time in order to identify the `optimal' set of parameters.

One may assume that to find the optimal set of parameters, where we are optimising over a single output metric -- referred to as an objective -- is just the process of running the simulation many times until we find the `best' set. Unfortunately, far too often, this is not the case. The search space over which parameters can vary and the number of possible parameters can be far larger than what can be feasibly (or economically) searched. One may conclude that each individual parameter may be optimised in isolation. However, if the relationship between parameters and the objective is complex, then the optimal value for one parameter may not be part of the global optimal. 

This complexity can be compounded when one wishes to optimise for multiple objectives, for example the energy used by a system and the increase in time to perform the work -- overhead. If one is fortunate, these objectives are mutually constructive and this degrades to a single optimisation case. However, in most cases multiple objectives are mutually destructive. In our example using only the lowest energy consuming computers could minimise energy consumption, though at the expense of delaying work completion when the number of available low-energy computers is insufficient.

In order to deal with optimising over multiple objectives one may choose to optimise for one objective over the other(s) or to optimise for a combination of them. However, this removes full transparency of the interplay between optimising for the different objectives -- diminishing the ability for decisions to be made from full knowledge.

We overcome the search space and multiple objectives problems by applying a Genetic Algorithm (GA)~\cite{mitchell1995genetic} to generate a Pareto frontier~\cite{Pareto1927, Stadler1979} using the Non-Dominated Sorting Genetic Algorithm II (NSGA-II)~\cite{Valkanas2014}. Using a GA allows us to quickly identify those parameters which lead to optimal output by selecting parameter sets which are mutations of the best sets identified in previous generations. Using NSGA-II allows us to identify those parameters which lead to objectives which lie along the Pareto frontier -- a curve which identifies those points for which there is no (yet identified\footnote{Note that as we do not try every combination of parameters it may be possible to improve these points.}) combination of parameters which would improve one of the objectives without diminishing the others.

We exemplify this for the digital twin tailored from the HTC-Sim~\cite{htc-sim} simulation system of a high-throughput computing (HTC) setup -- specifically HTCondor~\cite{htcondor} at Newcastle University. Our scheduler, which chooses which resources should run which tasks\footnote{Here a task, sometimes referred to as a job, is a single executable which is run on one computer within the HTCondor system.}, employs a Reinforcement Learning~\cite{rl} (RL) approach based on the work by McGough \textit{et al.}~\cite{suscom}. We identify twenty parameters from this work which can be used to configure the RL scheduler and consider the two objectives of energy consumed by the system and average task overhead -- difference between execution time and time in the system.

The rest of this paper is set out as follows. In Section \ref{motivation} we motivate the need for a GA along with Pareto frontier for the HTCondor RL scheduler. Related work is presented in Section \ref{relatedWork} followed by a discussion of the optimisation method in Section \ref{method}. Section \ref{environment} presents the simulation environment. Results are presented in Section \ref{results}. We offer conclusions and identify future directions in Section \ref{conc}.	
%
%
\section{Motivation}
\label{motivation}
Parameter spaces for simulations rapidly become large as the number of parameters and valid values increase. This is perhaps why in their work McGough \textit{et al.}~\cite{suscom} only ever vary two input parameters at a time and even then only consider a maximum of eleven different values for each of these parameters -- leading to 121 different simulations. 

Continuous value parameters are the hardest to deal with as selecting the size of discretisation is vitally important -- too small will lead to excessively large numbers of simulations, whilst too large means one is more likely to miss the optimal value. Integer values are similar in complexity but there is a minimum level of discretisation -- the unit value. 

Let us assume here that we wish to perform a parameter sweep over just six continuous values for a simulation which takes just five minutes per run. If we discretise each of the continuous parameters to one hundred values then we would require $10^{12}$ simulation runs, which is just over 9.5 million years of execution time. If we were to restrict the discretisation to just ten values per parameter this would reduce our parameter space to one million simulation runs and 9.5 years of execution time. Either case is far in excess of what can be performed -- and would be a significant energy drain in its own right. Thus the use of a machine learning optimisation approach is highly desirable here.

\begin{figure}[t]
\centering
\includegraphics[width=0.4\textwidth]{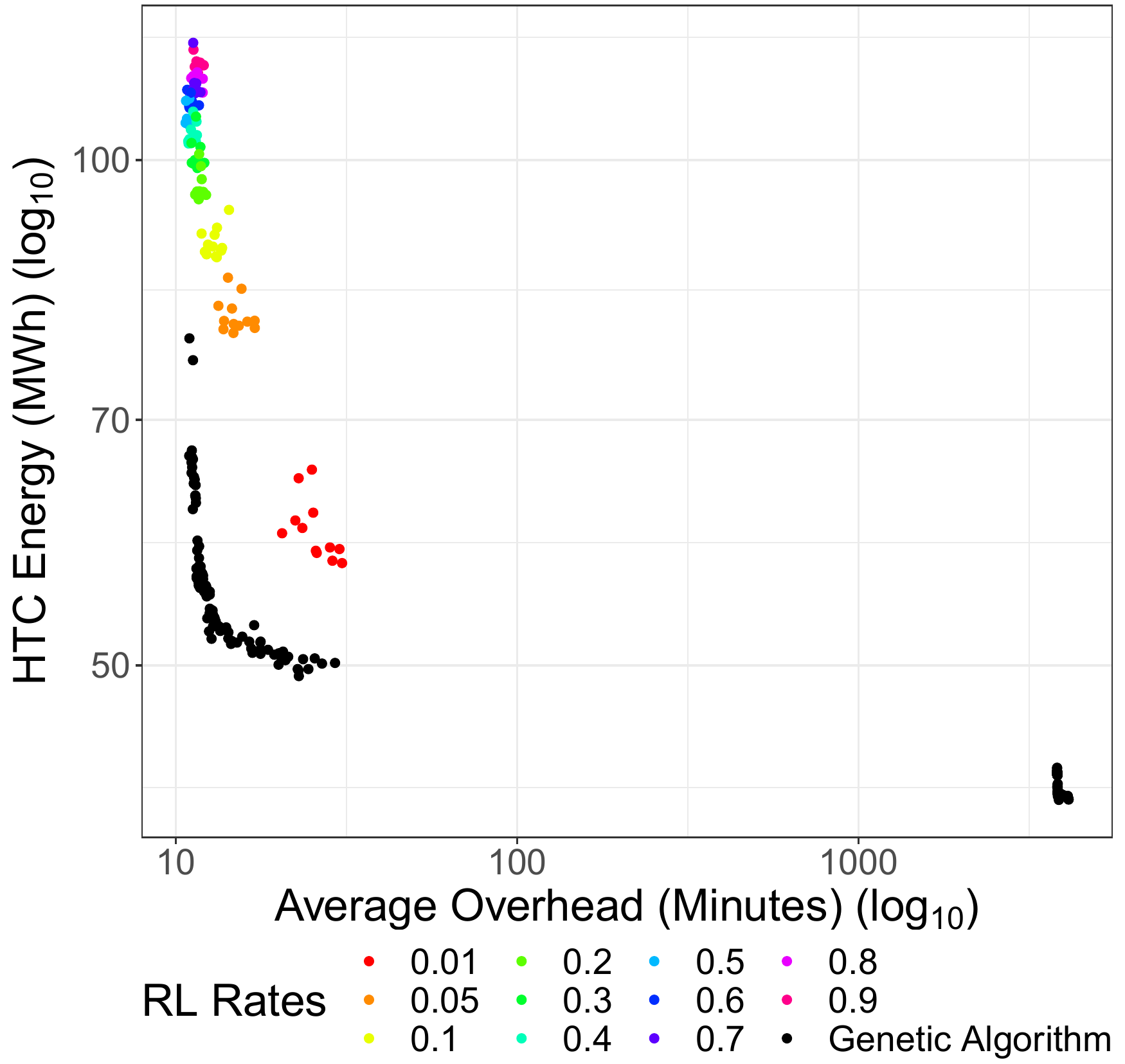}
\caption{Overhead vs Energy, adapted from \cite{suscom}, along with Pareto frontier}
\label{ohven}
\end{figure}

We use Figure \ref{ohven}, adapted from \cite{suscom}, to illustrate the motivation for identifying a Pareto frontier. The variation in colours represents the different learning rates ($\epsilon$) of the RL approach whilst the spread of each colour represents variations in how much importance is placed on the computers selected. It can be seen from this figure that there is no global optimal -- minimising energy and average overhead. This figure demonstrates that a Pareto frontier is present, though, due to the small number of sample points this is most likely not the actual Pareto frontier. For illustrative purposes we add here (black points) the identified Pareto frontier from our work. Demonstrating the savings which can be made by identifying more optimal parameter sets.
\section{Related Work}
\label{relatedWork}
Multi-objective optimisation problems are commonplace, with applications as diverse as electoral zone design \cite{Ponsich2017} to generation expansion planning \cite{Kannan2009}. Here we review various applications that have utilized multi-objective optimization.

Multi-objective optimization has been used in many different fields, and many multi-objective problems have been solved with Non-Dominated Sorting Genetic Algorithm II (NSGA-II) \cite{Valkanas2014}. There are, however, other algorithms which are used such as Multi-Objective Genetic Algorithm \cite{T.MurataandH.Ishibuchi1995}.

Ponsich \textit{et al.} apply NSGA-II to electoral zone design~\cite{Ponsich2017}. The criteria in which the geographical units must be aggregated are population equality, compactness and contiguity. They found that NSGA-II obtains promising results when compared with simulated annealing, producing better-distributed solutions over a wider-spread front.

Kannan \textit{et al.} used NSGA-II for the generation expansion planning problem  \cite{Kannan2009}. Seeking to identify which generating units should be commissioned and when they should become available over the long-term planning horizon. Optimising for two trade-off solutions: minimize cost, and minimize sum of normalized constraint violations; and to minimize investment cost and minimize outage cost. They were able to find a Pareto-front with high computational efficiency.

Wei \textit{et al.} used NSGA-II to optimize energy consumption and indoor environment thermal performance~\cite{Yu2015a}. Using simulation data containing energy consumption and indoor thermal comfort. They used a fitness function for NSGA-II comprising of a back propagation network optimised by a genetic algorithm to characterize building behaviour.

Guha \textit{et al.} used multi-objective optimization to design a ship hull \cite{Guha2015}. As their objective functions were not smooth, they found evolutionary techniques the most practical. They tested a number of different algorithms, and found that Sequential Quadratic Programming, Pattern Search and Interior-Point were very sensitive to the initial guess and prone to getting stuck in local minima. The genetic algorithm and particle swarm optimisation proved to be more robust and able to determine the global minima in most trials.

\section{Optimization methods}
\label{method}
Classical optimization methods, such as non-linear programming, find single solutions per simulation run. However, many real-world problems naturally have multiple objectives to optimise. Traditionally, optimization methods are used by converting them into a single-objective problem. However, this does not take into account the various trade-offs between equally optimal (Pareto-optimal) solutions. It is therefore important to find multiple Pareto-optimal solutions. A Pareto frontier is made up of many Pareto-optimal solutions. These can be displayed graphically, allowing a user to choose between various solutions and trade-offs.

Classical methods require multiple applications of an optimization algorithm, with various scalings between rewards to achieve a single reward. The population approach of genetic algorithms, however, enable the Pareto frontier to be found in relatively few simulation runs. NSGA-II is a multi-objective genetic algorithm and is used here.

\subsection{Genetic Algorithms}
GAs~\cite{Holland1975} are a class of evolutionary algorithms. We detail the workings of genetic algorithms in this section.

An initial population of structures $P_{0}$, for generation 0, is generated and each individual is evaluated for fitness. A subset of individuals, $C_{t+1} \subset P_{t}$, are chosen for mating, selected proportional to their fitness. `Fitter' individuals have a higher chance of reproducing to create the offspring group $C'_{t+1}$. $C'_{t+1}$ have characteristics dependent on the genetic operators: crossover and mutation. The genetic operators are an implementation decision \cite{FogelDavidB2009}. 

Once the new population has been created, the new population $P_{t+1}$ is created by merging individuals from $C'_{t+1}$ and $P_{t}$. See Algorithm \ref{genetic-algorithm} for detailed pseudocode.
\begin{algorithm}[t]
\begin{algorithmic}[1]
\State $t=0$
\State initialize $P_{t}$
\State evaluate structures in $P_{t}$
\While {termination condition not satisfied}
\State $t=t+1$
\State select reproduction $C_{t}$ from $P_{t-1}$
\State recombine and mutate structures in $C_{t}$

forming $C'_{t}$
\State evaluate structures in $C'_{t}$
\State select each individual for $P_{t}$ from $C'_{t}$ 

or $P_{t-1}$
\EndWhile
\caption{Genetic algorithm \cite{FogelDavidB2009}}
\label{genetic-algorithm}
\end{algorithmic}
\end{algorithm}
\subsection{NSGA-II}
NSGA-II is efficient for multi-objective optimization on a number of benchmark problems and finds a better spread of solutions than Pareto Archived Evolution Strategy (PAES)~\cite{Knowles1999} and Strength Pareto EA (SPEA)~\cite{Zitzler2006} when approximating the true Pareto-optimal front \cite{Valkanas2014}.

The majority of multi-objective optimization algorithms use the concept of \emph{domination} during population selection \cite{Burke2014}. A non-dominated genetic algorithm seeks to achieve the Pareto-optimal solution, so no single optimization solution should dominate another. An individual solution $\mathbf{x}^{1}$ is said to dominate another $\mathbf{x}^{2}$, if and only if there is no objective of $\mathbf{x}^{1}$ that is worse than objective of $\mathbf{x}^{2}$ and at least one objective of $\mathbf{x}^{1}$ is better than the same objective of $\mathbf{x}^{2}$ \cite{Bao2017}. Non-domination sorting is the process of finding a set of solutions which do not dominate each other and make up the Pareto front. A Pareto front contains solutions that have dominated all inferior solutions, and have at least one objective which is better than the other solutions of the Pareto front. See Figure \ref{fig:pareto-layering}a for a visual representation, where $f_1$ and $f_2$ are two objectives to minimise.

We define a process to determine which solutions to keep:
\subsubsection{Non-dominated sorting}
We assume that there are $M$ objective functions to minimise, and that ${\bf x^{1}} = x_j^{1}$ and $\bf x^{2}$ are two solutions. $x_j^{1}<x_j^{2}$ implies solution $\bf x^{1}$ is better than solution $\bf x^{2}$ on objective $j$. A solution $\mathbf{x}^{1}$ is said to dominate the solution $\mathbf{x}^{2}$ if the following conditions are true:
\begin{enumerate}
  \item The solution $\mathbf{x}^{1}$ is no worse than $\mathbf{x}^{2}$ in every objective. I.e. $x^{1}_j \leq x^{2}_j \;\;  \forall j \in\{1,2,\ldots,M\}$.
  \item The solution $\mathbf{x}^{1}$ is better than $\mathbf{x}^{2}$ in at least one objective. I.e. $\exists\  {j}\in \{ 1,2,\ldots,M\} \;\; s.t. \;\;x^{1}_j < x^{2}_j$.
\end{enumerate}

\begin{figure}[t] 
  \center
  \includegraphics[width=0.2\textwidth]{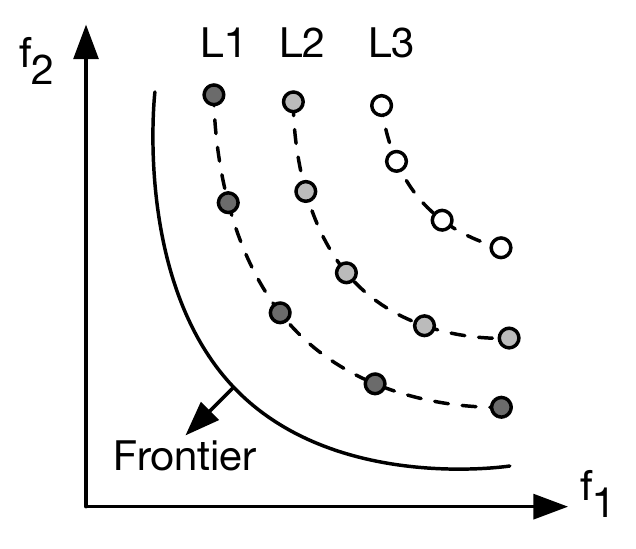}
  \includegraphics[width=0.275\textwidth]{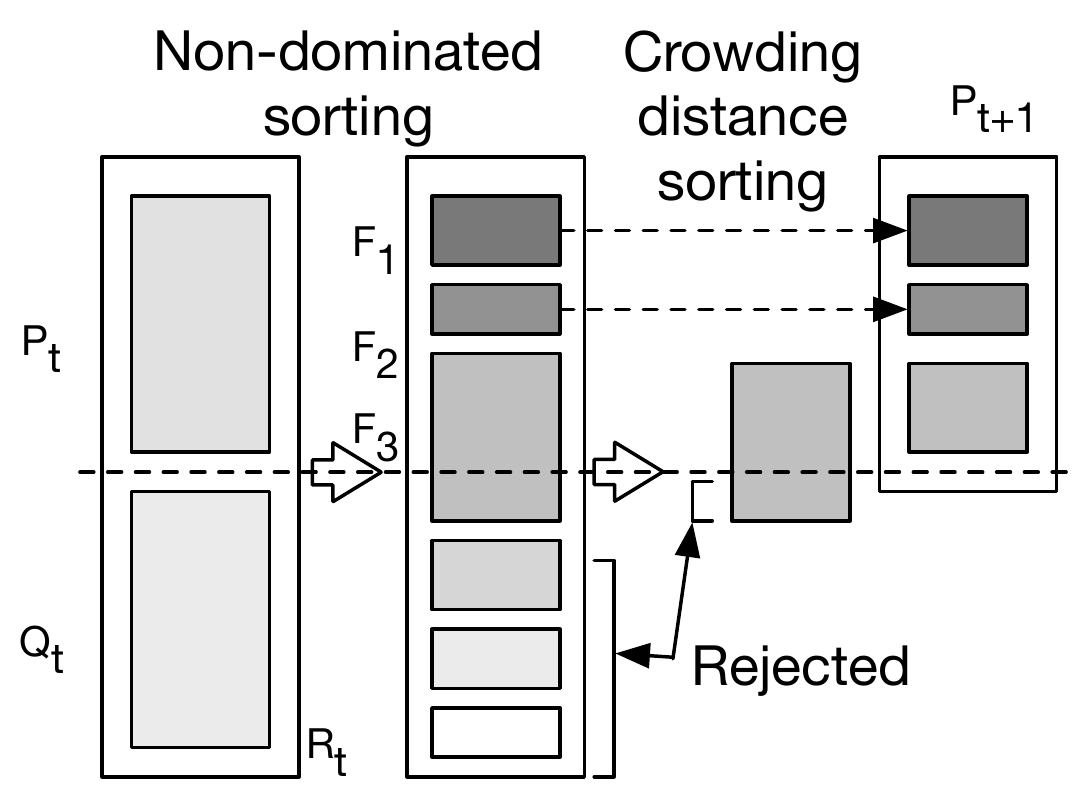}
  \caption{a) Schematic of non-dominated sorting with solution layering b) Schematic of the NSGA-II procedure}
  \label{fig:pareto-layering}
\end{figure}

Once the solutions are calculated for the $M$ objective functions, the solutions are sorted according to their level of non-domination. An example of layering of levels is shown in Figure \ref{fig:pareto-layering}a. Here, $f_1$ and $f_2$ are the objective functions to be minimized. The Pareto front is the first front which contains solutions that are not dominated by any other solution. The solutions in layer 1 are dominated only by those in the Pareto front, and are non-dominated by layer 2 and layer 3. 

The solutions are then ranked according to their layer. Solutions in the Pareto front are given a fitness rank ($i_{rank}$) of 1, solutions in layer 1 have $i_{rank}$ of 2, etc.

\subsubsection{Density Estimation}
($i_{distance}$) is computed for each solution as the average distance between the two closest points to the solution in question, and is an estimate of the largest cuboid which contains only $i$ and no other points. 

\subsubsection{Crowded comparison operator}
($\prec_n$) is used to ensure that the final frontier is an evenly spread out Pareto-optimal front. Each solution has two attributes: $(i_{rank})$ and$(i_{distance})$. 
We can then define a partial order:\\	
$i\prec_nj$ if $(i_{rank}<j_{rank})$ or $((i_{rank}=j_{rank})$ and  $(i_{distance}>j_{distance}))$ \cite{Valkanas2014}.

This concludes that a point with a lower rank is preferred, and if two points have the same rank the point which is located in a less dense area is preferred.

\subsubsection{Main loop}
As with standard GA a random population $P_{0}$ is created. This is then  sorted according to non-domination. Binary tournament selection, recombination and mutation operators are used to create a child population $C'_{1}$ of size $N$. Where tournament selection is a process of evaluating and comparing the fitness of various individuals in a population. Binary tournament selection begins by selecting two individuals at random, evaluating the fitnesses, and selecting the individual with the better solution \cite{AbdRahman2016}.

\begin{algorithm}[b]
\begin{algorithmic}[1]
\State $R_t=P_t \cup C'_t$ combine parent and child population
\State $\mathcal{F} = $ fast-non-dominated-sort $(R_t)$ 

where $\mathcal{F}=(\mathcal{F}_1, \mathcal{F}_2,\ldots)$
\State $P_{t+1}=\emptyset$
\While $\left|P_{t+1}<N\right|$
\State Calculate the crowding distance of $(\mathcal{F}_i)$)
\State $P_{t+1}=P_{t+1}\cup \mathcal{F}_i$
\EndWhile
\State Sort($P_{t+1}, \prec_n$) sort in descending order using $\prec_n$
\State $P_{t+1} = P_{t+1}[0:N]$ select the first $N$ elements of $P_{t+1}$
\State $Q_{t+1} = $ make-new-population$(P_{t+1})$ using 

selection, crossover and mutation to create 

the new population $Q_{t+1}$
\State $t=t+1$
\caption{NSGA-II main loop \cite{Valkanas2014}}
\label{algo:nsga2}
\end{algorithmic}
\end{algorithm}
After the first population the procedure changes (see Algorithm \ref{algo:nsga2}). Initially, a combined population is formed $R_{t}=P_{t} \cup C'_{t}$ of size $2N$. $R_{t}$ is sorted according to non-domination. A new population is now formed $(P_{t+1})$, adding solutions from each front level until the size of $P_{t+1}$ exceeds $N$. The solutions of the last accepted level are then sorted according to $\prec_n$, and a total of $N$ solutions are chosen, rejecting those from the last layer that have a smaller crowding distance \cite{Valkanas2014}.

The entire process is shown in Figure \ref{fig:pareto-layering}b, and is repeated until the termination condition is met. Termination conditions could be: no significant improvement over $X$ iterations or a specified number of iterations have been performed.

%

\section{Simulation environment}
\label{environment}
\subsection{HTC-Sim}
HTC-Sim is a trace-driven simulation framework for energy consumption in High Throughput Computing systems~\cite{htc-sim}. The simulation handles two types of users -- interactive users who can sit down in-front of a computer and use it along with high-throughput users who submit multiple tasks through a batch submission system which use the computers when idle. Interactive users will evict HTC tasks requiring the task to be rerun. The computers in the system are considered at three logical levels -- the whole system, a cluster of computers (a number of computers in a distinct location) and individual computers.

The model characterises each computer through a set of parameters. These describe the resource in terms of operating system, architecture type, memory size, performance metrics (such as number of cores, CPU speed, MIPS), along with an energy profile. The model is extensible, allowing practitioners to define their own custom parameters.

The workload of the HTC system is comprised of a set of high throughput tasks,  submitted either independently or together as part of a batch. A task submitted to the system is initially placed into a queue. If an appropriate computer is available the task is allocated to that resource -- the task is now in the running state. If no appropriate computer is available the task will remain queued until an appropriate computer is available. If an interactive user logs into the computer whilst a high throughput task is running, the task will relinquish the resource either by entering a suspended state (if possible) or re-entering the queue to be re-run later. Tasks that remain in a suspended state for longer than a pre-determined threshold are evicted and re-enter the queue.

An ordered set of all interactive sessions is used to replay the interactive user activity across the computers within an organisation. The data used to exemplify the system is trace data obtained from December 2009 through December 2010. These traces are indicative of current system usage and analysis that has been ongoing since 2010.

We are primarily concerned with two objectives (metrics):

\textbf{Average task overhead} -- the time difference between the task entering ($q_t$) and departing ($f_t$) the system, and the actual task execution time ($d_t$) for a set of tasks $T$:
\begin{equation*}
1/|T|\sideset{}{_{t \in T}}\sum (f_{t} - q_{t} - d_{t}).
\end{equation*}
\textbf{Energy consumption} -- the total energy consumed by the HTC workload. Fine-grained energy consumption is recorded per- computer, cluster and system, for each state, e.g. sleep, idle, active (HTC and/or interactive user). The total energy consumption is then calculated as follows:
\begin{equation*} 
\sideset{}{^n_{c=0}}\sum \sideset{}{^m_{p=0}}\sum t_{c,p} E_{c,p},
\end{equation*}
\noindent where $n$ is the number of computers, $m$ is the number of power states, $t_{c,p}$ is the time spent by computer $c$ in state $p$ and $E_{c,p}$ is the power consumption rate of computer $c$ in state $p$. For non-HTC states, $E_{c,p} = 0$.
\subsection{Reinforcement Learning Scheduler}
Reinforcement Learning~\cite{rl} (RL) is a machine learning technique used to learn how to react to an environment. An agent observes an environment which is often represented by a state space. For each state in the state space there is a corresponding action vector representing every action which can be taken in that state. Initially each action has the same probability of being selected. When the agent observes a specific state it chooses an action from the action vector based on either an explorative or exploitative policy -- selected between at random with probability $\epsilon$. If an explorative policy is chosen then the action is selected at random from the action vector whilst if an exploitative policy is in force then the action which has seen the `best' historical reward is selected. Once the action is completed and it is known if the action was good or bad then the reward value for the action is updated -- rewarding good actions (increasing the reward) and punishing bad actions (decreasing the reward).

The RL scheduler by McGough {\em et al.}~\cite{suscom} has a state space which is a combination of whether computers within the HTC system are free for use and the hour of the day when the request to schedule a task is made. The granularity of the action space can be varied in size from representing each computer individually through to only selecting the cluster on which to place a task or placing in the queue, to only selecting between allocating the tasks to a computer or queueing the task. Likewise the hour of the day could be for any day (24 actions) or for each hour within a week (168 actions). 


\subsection{Parameters for RL}
We present here the parameters that the GA will search over in order to identify the optimal policies. Further details can be found in \cite{suscom}.

The exploration versus exploitation of a RL approach is potentially the most significant factor in optimising the approach. Too small a value of $\epsilon$ will lead to the system not searching the possible outcome space and hence performing little better than randomly choosing actions. Likewise, too large a value of $\epsilon$ will mean the RL is spending more time searching for optimal solutions than actually using the ones that it has found already. However, having a single value of $\epsilon$ for the whole RL process can be too restrictive. We therefore allow $\epsilon$ to decrease as the simulation progresses. Below we present parameters which control the action space, the reward computation and  how the value of $\epsilon$ is varied:
\begin{itemize}
	\item {\bf Week}: is the state space for the RL -- day or week {\em \{boolean\}}. As weekends have a different usage patten to week days this could allow the RL to adapt to this.
	\item {\bf Entity level}: is the component of action space in terms of computer granularity -- {\em \{(computer, cluster, whole)\}}. Fine grained actions could be better, but at the expense of needing far more examples to train on. 
	\item {\bf $\epsilon$-policy}: What is $\epsilon$ changed on? {\em \{(days, previous, ratio, hit)\}}. Note that this has an impact on the meaning of many of the parameters below.
	\item {\bf ranges}: The date range on which to change $\epsilon$. {\em \{[0,999999], ...\}}\footnote{Although this can be an arbitrarily long list we limit this to three values for this work.},\footnote{Note $value_i \leq value_{i+1}$}.
	\item {\bf reward boundaries}: reward values over which the $\epsilon$ value will be changed. {\em \{(0,1], ...\}}$^3$,$^4$
	\item {\bf $\sigma$}: The amount of influence the computer energy efficiency has on RL reward {\em \{[0,1]\}}.
	\item {\bf days}: Change $\epsilon$ based on the number of days of RL which have been performed {\em \{[0, 365]\}, 0 = don't use.}
	\item {\bf $\delta$}: Increase $\epsilon$ by $\delta$ if current day into RL	$\leq$ days {\em \{[0,1]\}}.
	\item {\bf history}: The number of previous tasks to consider when computing the action {\em \{[-1,999999]\}}, -1 = all tasks.
	\item {\bf gaussian}: Do we apply a gaussian decay over the task history when computing the action? {\em \{boolean\}}.
	\item {\bf prior}: If all prior actions gave a negative reward then increase $\epsilon$ by 0.1 {\em \{boolean\}}.
	\item {\bf threshold}: If the ratio of best reward to average reward is less than threshold, use the previous $\epsilon$ value{\em \{[0,1]\}}.
	\item {\bf defer}: Defer running a task during the same hour that the computer is due to be rebooted {\em \{boolean\}}.
\end{itemize}
                                                                
If we assume here, conservatively, that each continuous parameter is discretised into one hundred values then this creates a search space of some $2.81 \times 10^{37}$. Again, assuming that the simulation takes five minutes to run for each parameter combination this is $2.67 \times 10^{32}$ years for a full parameter space search.

\section{Results}
\label{results}
\begin{figure}[t]
\centering
  \includegraphics[width=0.9\linewidth]{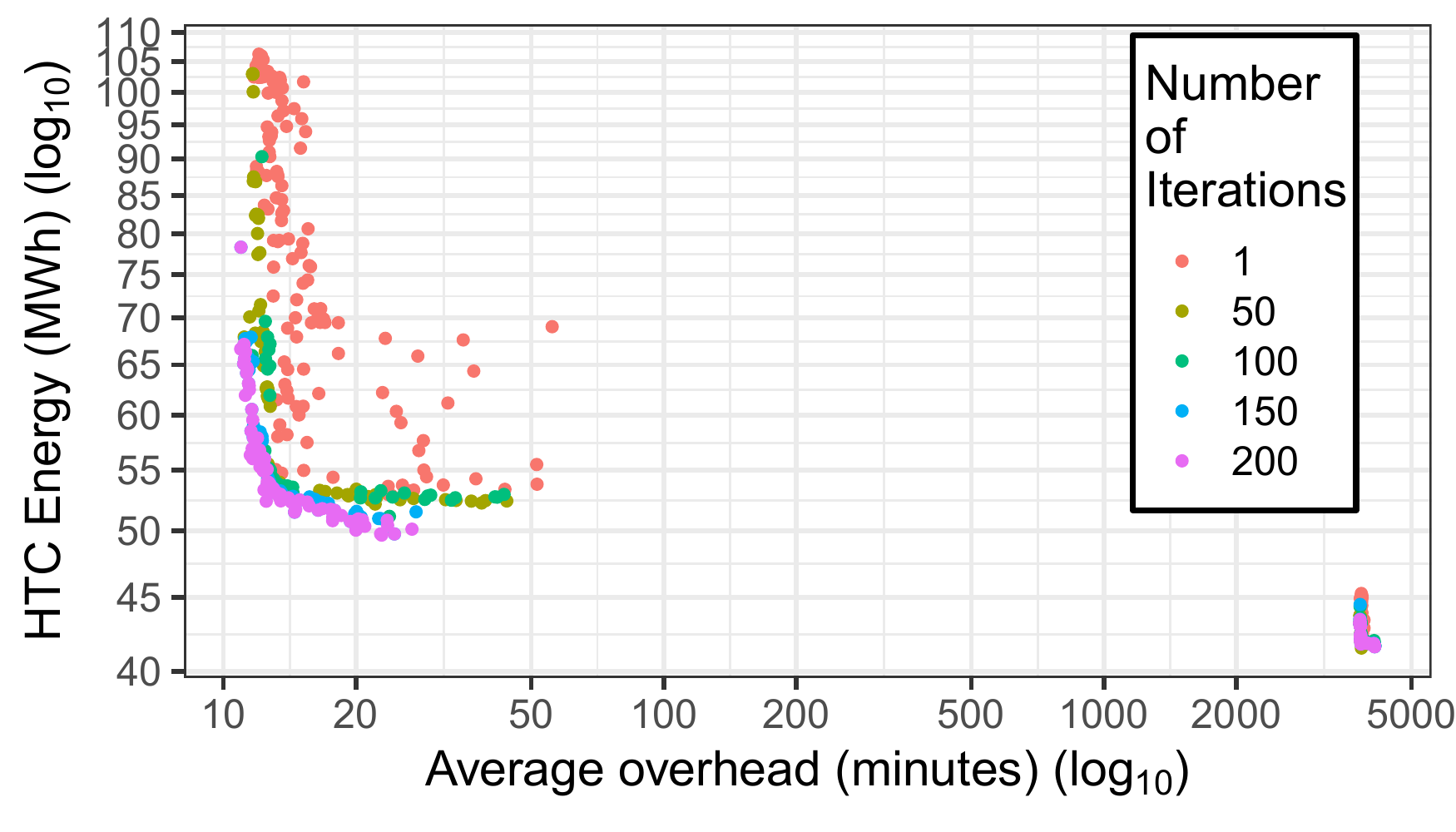}
  \caption{Progress towards the Pareto Frontier}
  \label{fig:pareto}
\end{figure}

\begin{table*}[tb]
\caption{Parameters for optimal objectives}
\label{tab:optimal}
\centering
\includegraphics[width=1\linewidth]{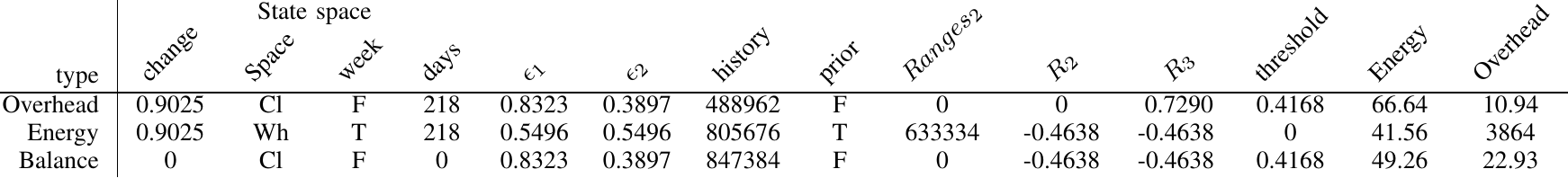}
\end{table*}
We first evaluate if the NSGA-II approach will lead to a Pareto frontier for total energy consumed and average overhead. Figure \ref{fig:pareto} illustrates progressive iterations of the NSGA-II algorithm with successive iterations in different colours. The initial (red) colours are scattered widely whilst the final iteration (magenta) indicates a sharp edge closest to the two axis. It is interesting to note that although there is no global optimal for both energy and overhead the Pareto front is `sharp' in the bottom left corner indicating that there is a good compromise for both objectives. There is also a separate region of points with lower energy consumption but substantially higher overheads. This appears to be cases where tasks are only allowed to run when the task is almost definitely going to finish -- at the expense of significantly increasing overhead. For all parameter sets along the Pareto front {\em defer} was true, demonstrating that not running tasks during the hour when a computer will be rebooted was the best policy. We therefore consider {\em defer} no further.

Table \ref{tab:optimal} presents the parameter sets and objective values for minimum overhead, minimum energy and `optimal' combination of overhead and energy. The optimal combination was attained by first scaling average overhead and total power consumed between 0 and 100 using min-max scaling. Next, we summed the scaled overhead and total power consumed and chose the combination with the minimum value. This enabled us to choose the minimum combination of both objectives with equal weighting. The scaling, however, could be changed to suit individual preferences of power consumption or average overhead. As the following parameters were identical for all cases, we present them here rather than in the table: $\epsilon$-policy was previous, gaussian was false, $ranges_1$ was 0, ratio was 0.9375 and $R_1$ was -0.5842.

The energy consumption here is far better than those presented in the paper by McGough {\em et al.}~\cite{suscom}, reducing the energy consumption by ${\sim}39$MWh for effectively the same average overhead whilst also being able to beat their best energy case by ${\sim}8$MWh again for no appreciative change in average overhead. We can save over $16$MWh of energy over their lowest energy case, however, this is at the cost of massively increasing the overhead. It should be noted that this was achieved solely through the tuning of the simulation parameters with the use of NSGA-II, as both sets of results run the same underlying code.

\begin{figure}[!b]
\centering
\begin{subfigure}{}
\centering
\includegraphics[width=0.48\linewidth]{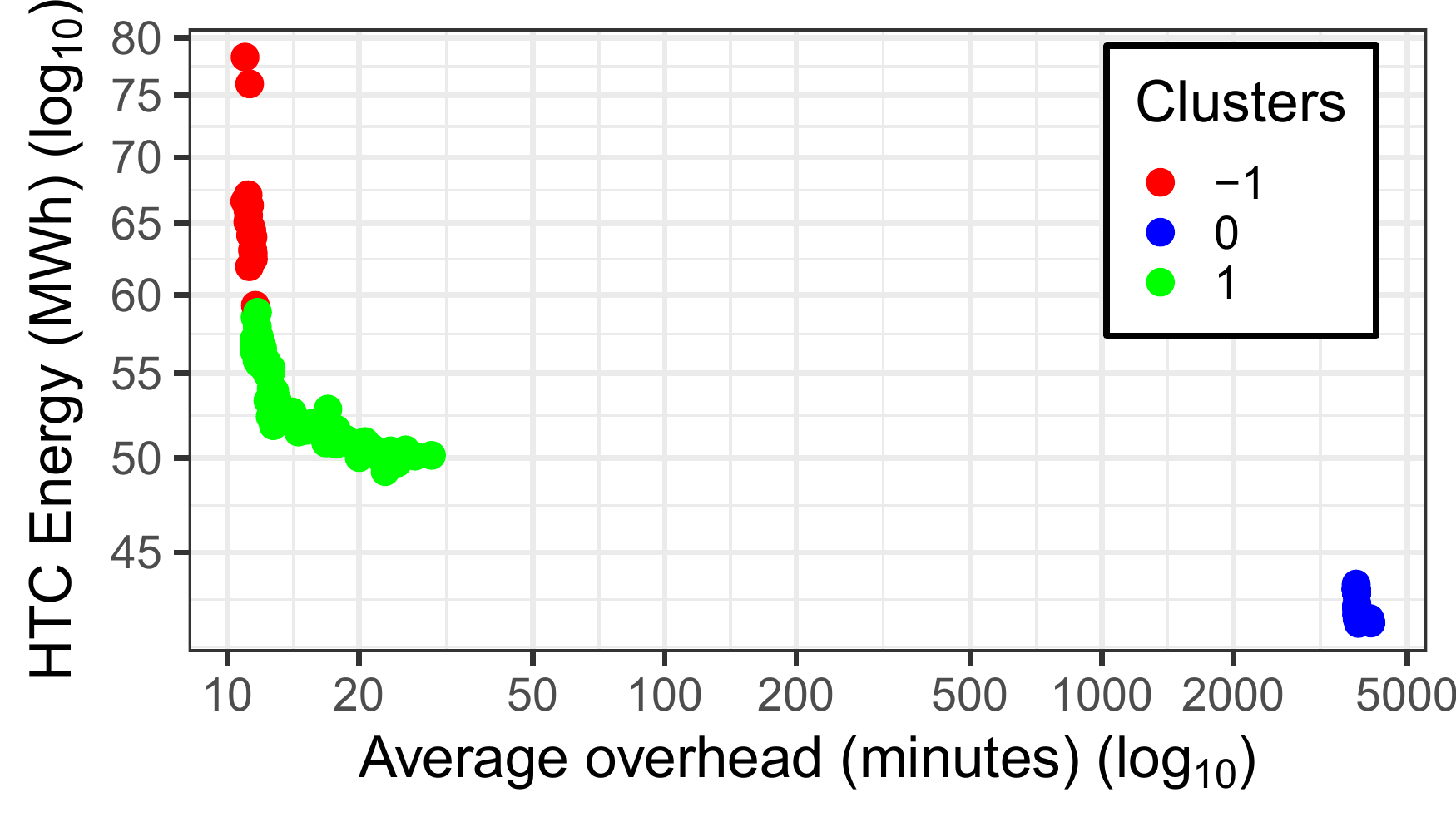}%
\hfil
\end{subfigure}
\begin{subfigure}{}
\centering
\includegraphics[width=0.48\linewidth]{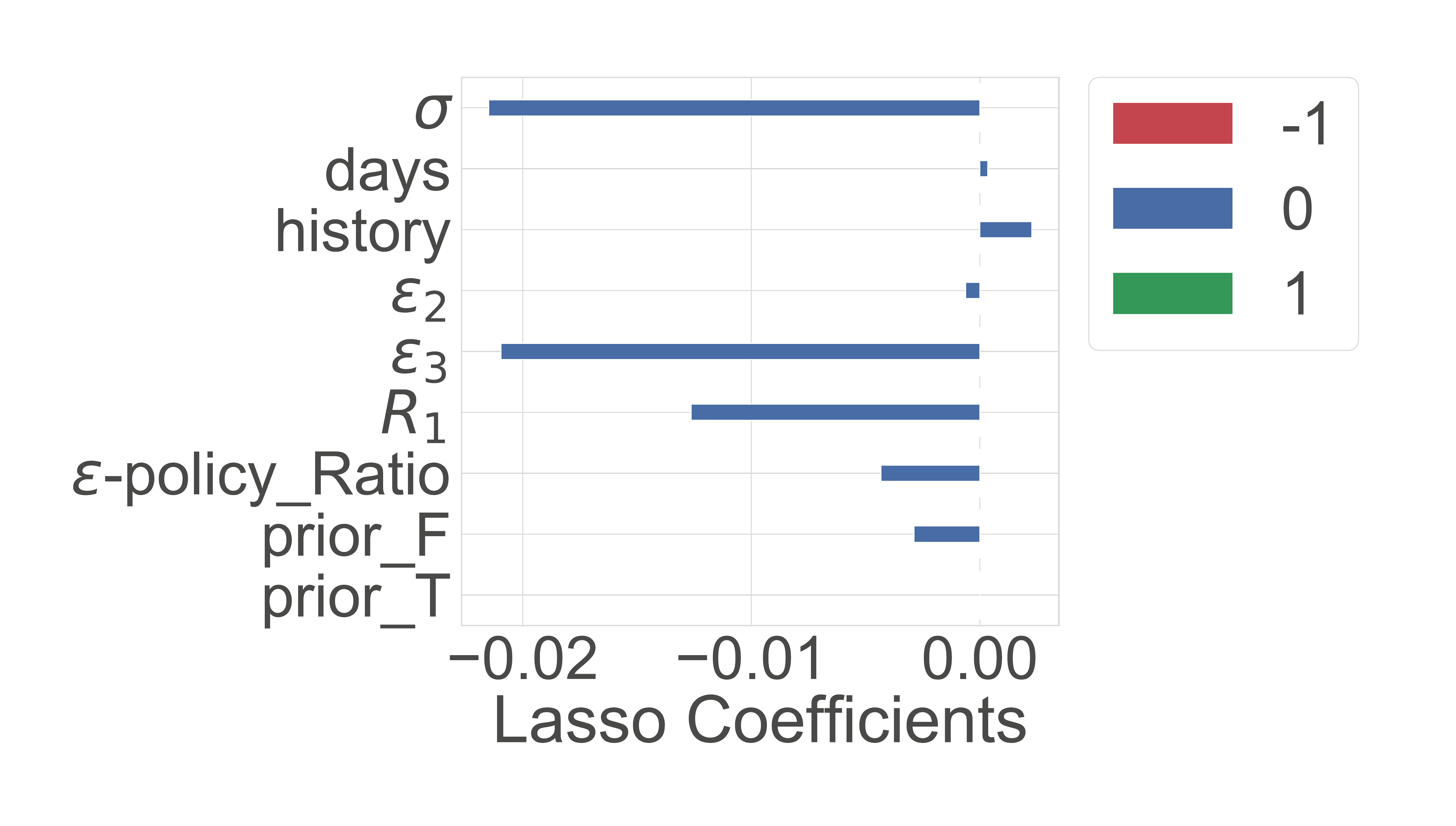}%
\hfil
\end{subfigure}
%
\caption{Objective clustering a) Clusters, b) Overhead impact}
\label{fig:clustering}
\end{figure}

To better understand the parameters which effect the Pareto Front we fit a Lasso regression \cite{Tibshirani1996} to distinct clusters of the Pareto Front (optimisation dominant (-1), central (1) and energy dominant (0) -- Figure \ref{fig:clustering}a). Lasso regression is a linear regression technique which steers the coefficients for insignificant parameters to zero allowing for the identification of important parameters and their significance. The parameters were all scaled between 1 and 100 allowing direct comparison between parameters.

We clustered the data using the unsupervised learning technique DBSCAN \cite{Ester:1996:DAD:3001460.3001507}. This technique was chosen due to its effectiveness at clustering data points which are close together. This yields better results for our dataset than a method such as k-means clustering which partitions the data into Voronoi cells. The results, Figure \ref{fig:clustering}b, show a large negative coefficient for $\sigma$, $\epsilon_3$ and $R_1$ when predicting average overhead, though only for the significant overhead case (cluster 0). This would somewhat suggest that looking at the energy efficiency of computers in these cases is detrimental -- potentially as this cluster favours queueing tasks rather than running them.

\begin{figure}[t]
\includegraphics[width=0.95\linewidth]{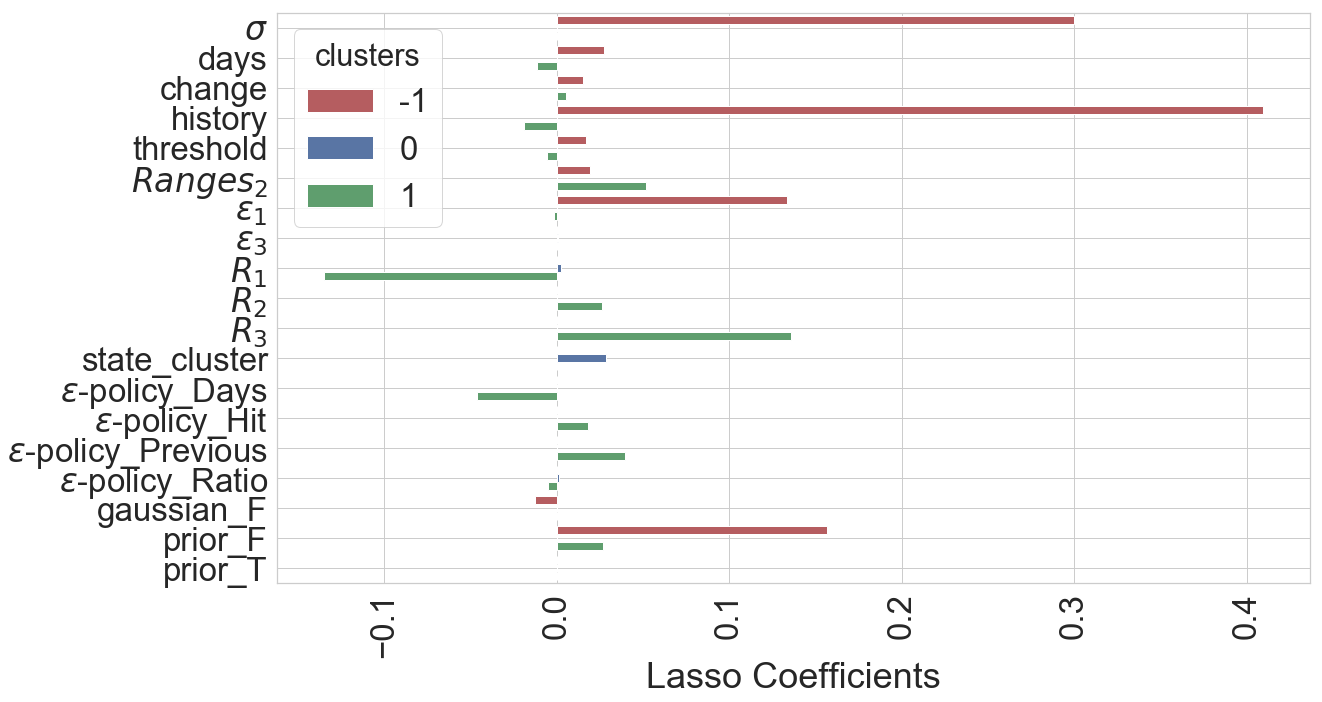}%
\caption{Parameters which impact energy consumption}
\label{fig:impact-energy}
\end{figure}

The coefficients for total energy, Figure \ref{fig:impact-energy}, show that lower values of $\sigma$ and history have the best impact on reducing energy for cluster -1 -- low overhead cases. This is against the naive assumption that taking energy efficiency into account would reduce overall energy consumption -- potentially as this could reduce clarity for which computer to use. Shorter history would suggest that the system changes over time and hence only recent history should be considered.

\begin{figure}[!t] 
\centering
\includegraphics[width=0.7\linewidth]{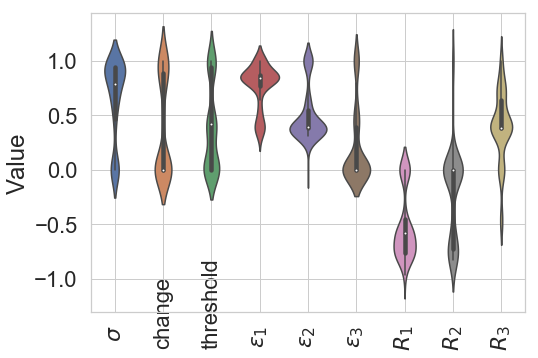}%
\caption{Violin distribution of parameters}
\label{fig:violin-parameters}
\end{figure}

Figure \ref{fig:violin-parameters} displays the distribution of parameters for $\varepsilon$, $R$, $\sigma$, change and threshold for the final population. It can be seen that for $\varepsilon$, the parameters converge to a high value for $\varepsilon_1$, low value for $\varepsilon_3$ with $\varepsilon_2$ between these two values. This is to be expected as the reward boundaries should decrease. $\sigma$ and $change$ are both bimodal, whereas $threshold$ is trimodal. $R$ converges to an increasing relationship with successive $R$, however, it is less defined than $\varepsilon$, with a bimodal relationship for $R_1$ and $R_2$.
\begin{figure}[t]
%
 \includegraphics[width=0.5\linewidth]{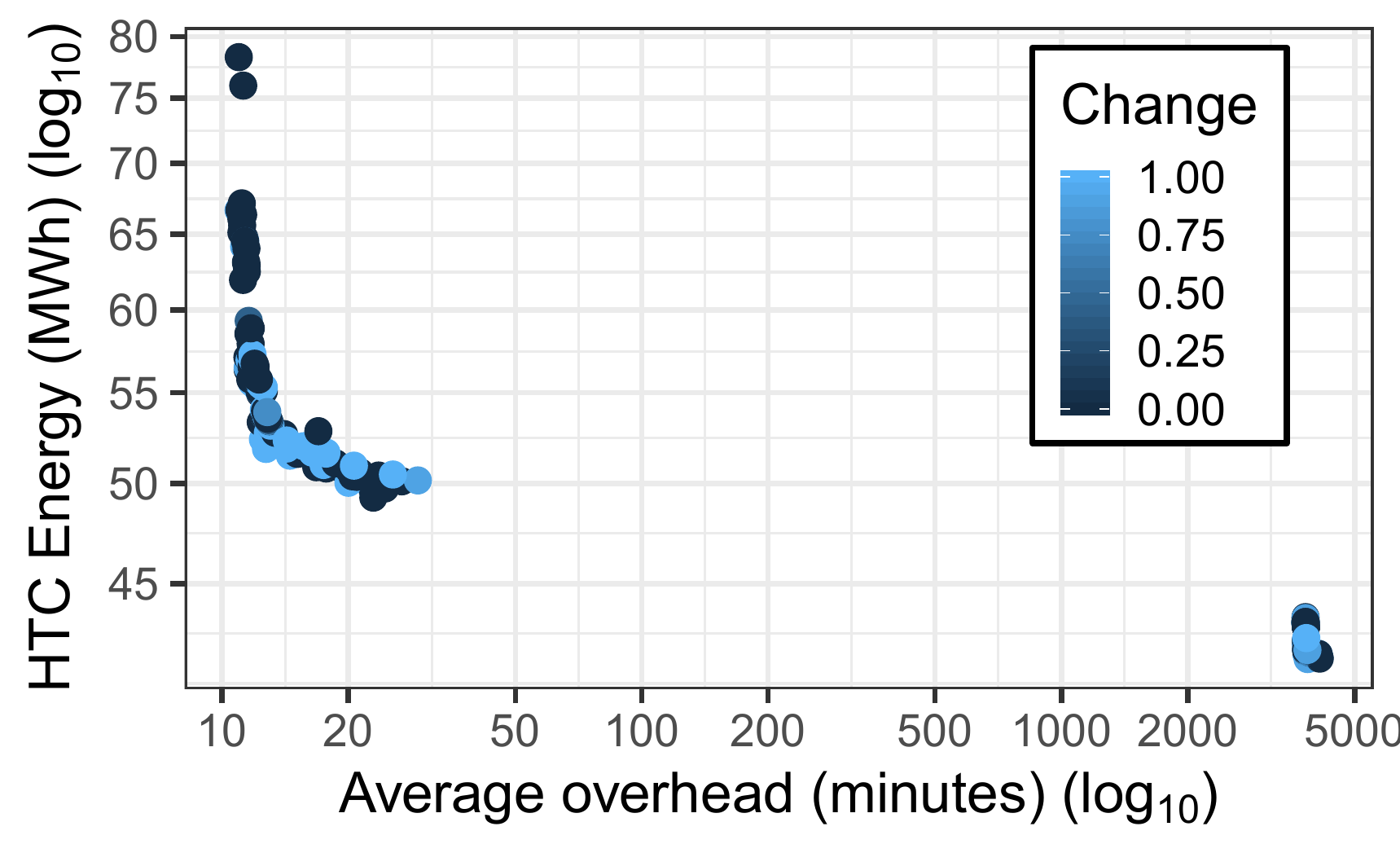}%
\hfil%
\includegraphics[width=0.5\linewidth]{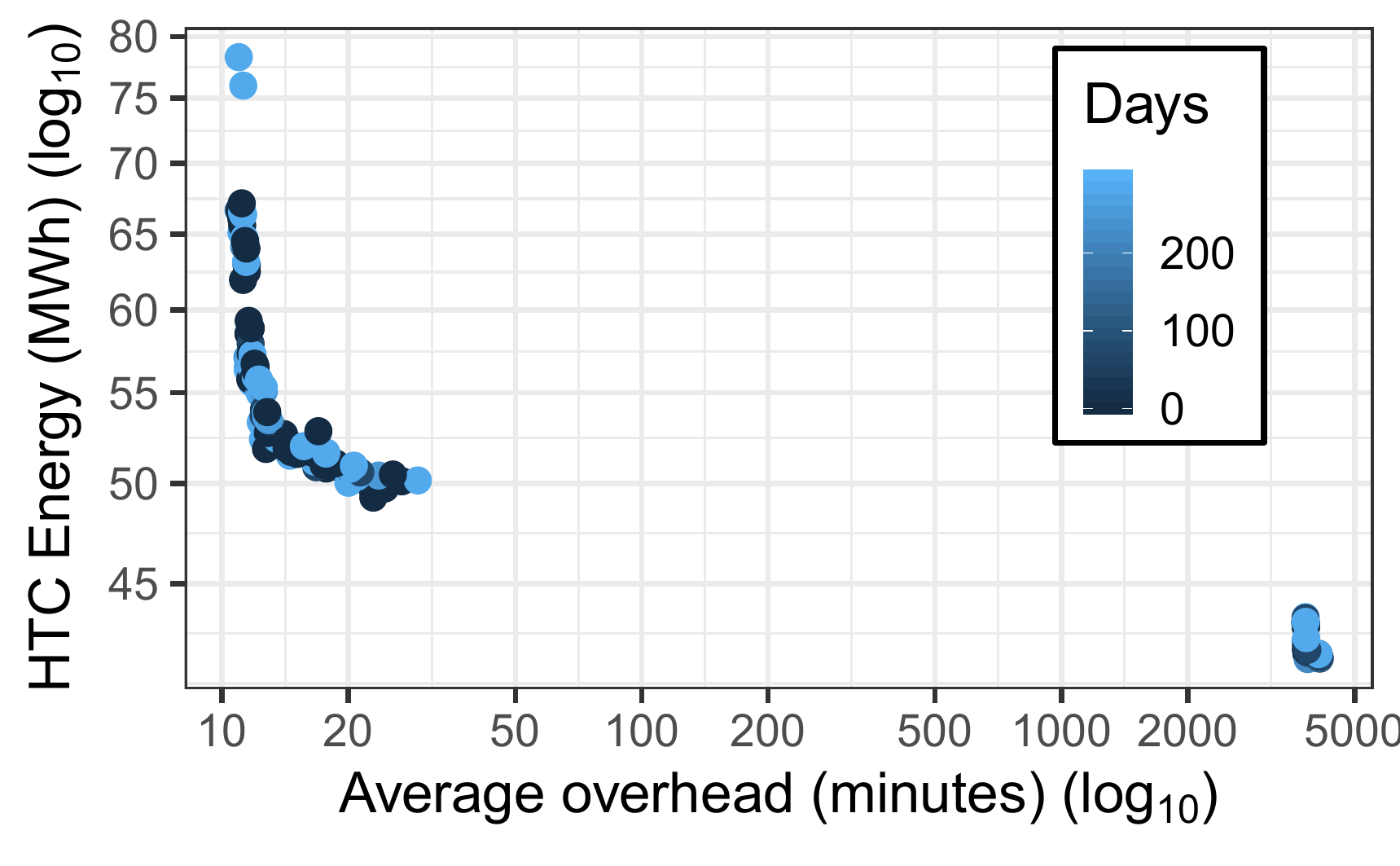}%
%
\caption{$\delta$ value and days impact on $\epsilon$}
\label{fig:changes}
\end{figure}
\begin{figure}[!t]
  \includegraphics[width=0.49\linewidth]{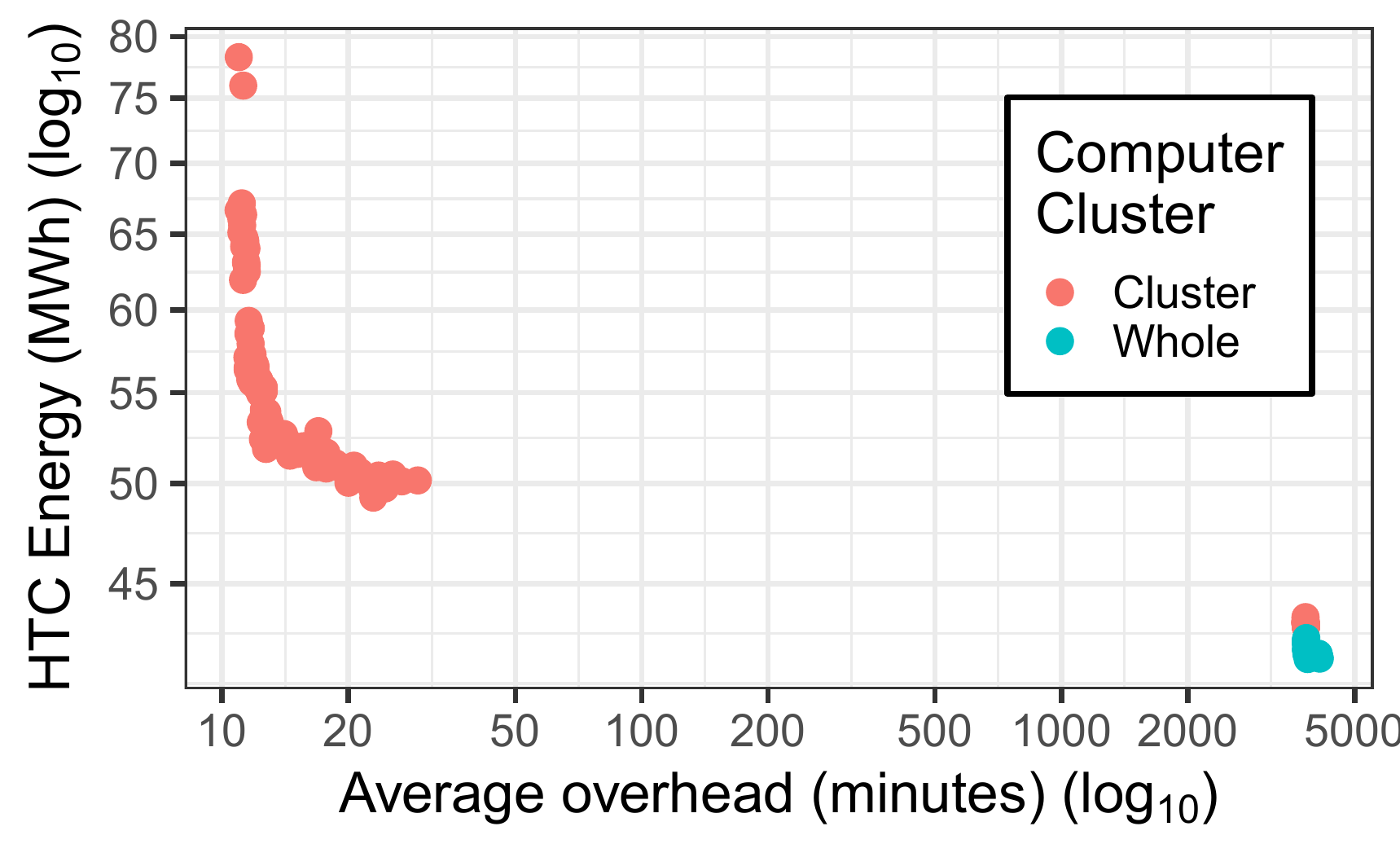}
  \includegraphics[width=0.49\linewidth]{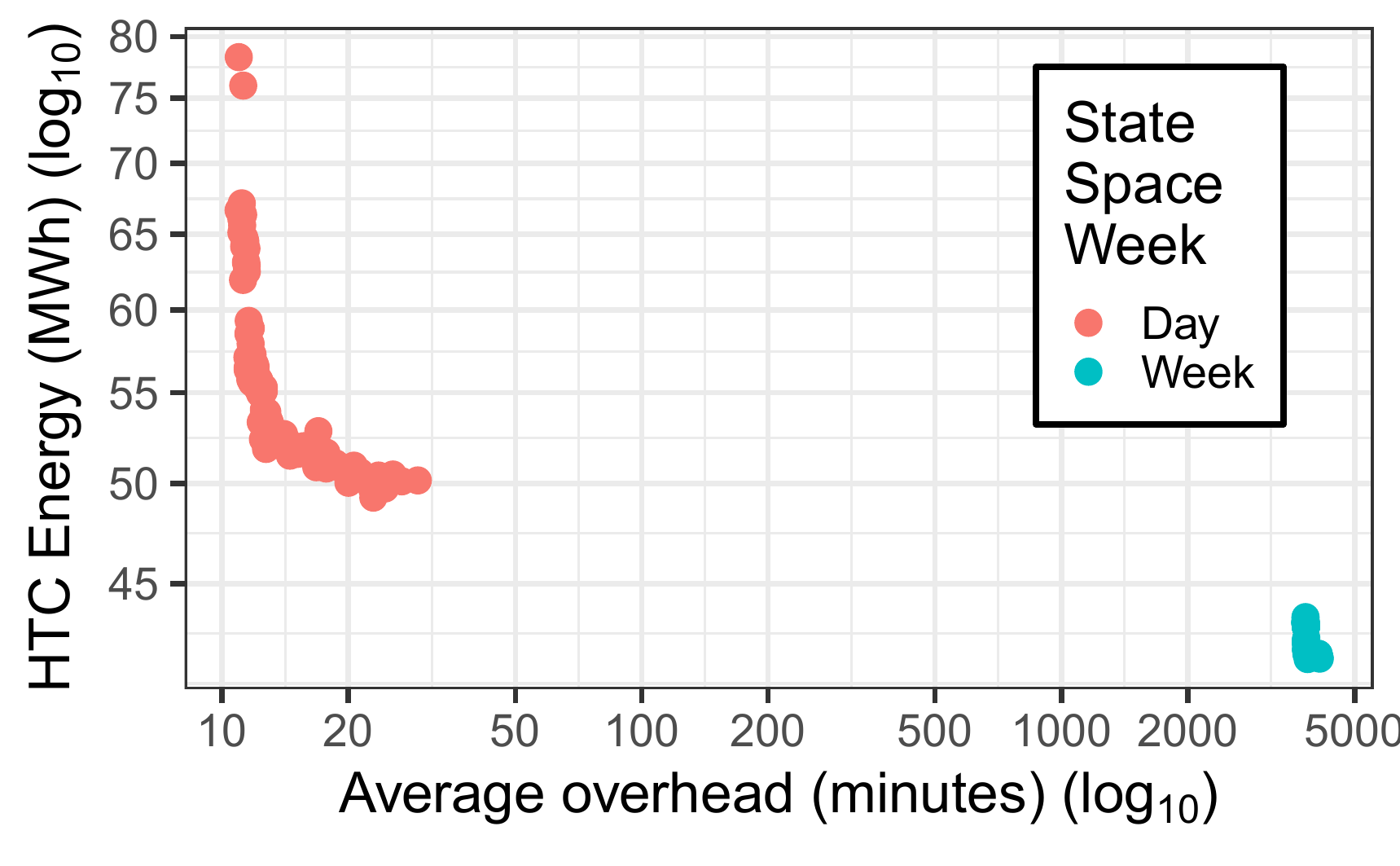}
  \caption{a) Granularity of RL action space with respect to computer- and cluster-level. b) Day / Week granularity for action/state space.}
  \label{fig:granularity}
\end{figure}
%
%
%
\begin{figure*}[!t]
  \includegraphics[width=0.3\linewidth]{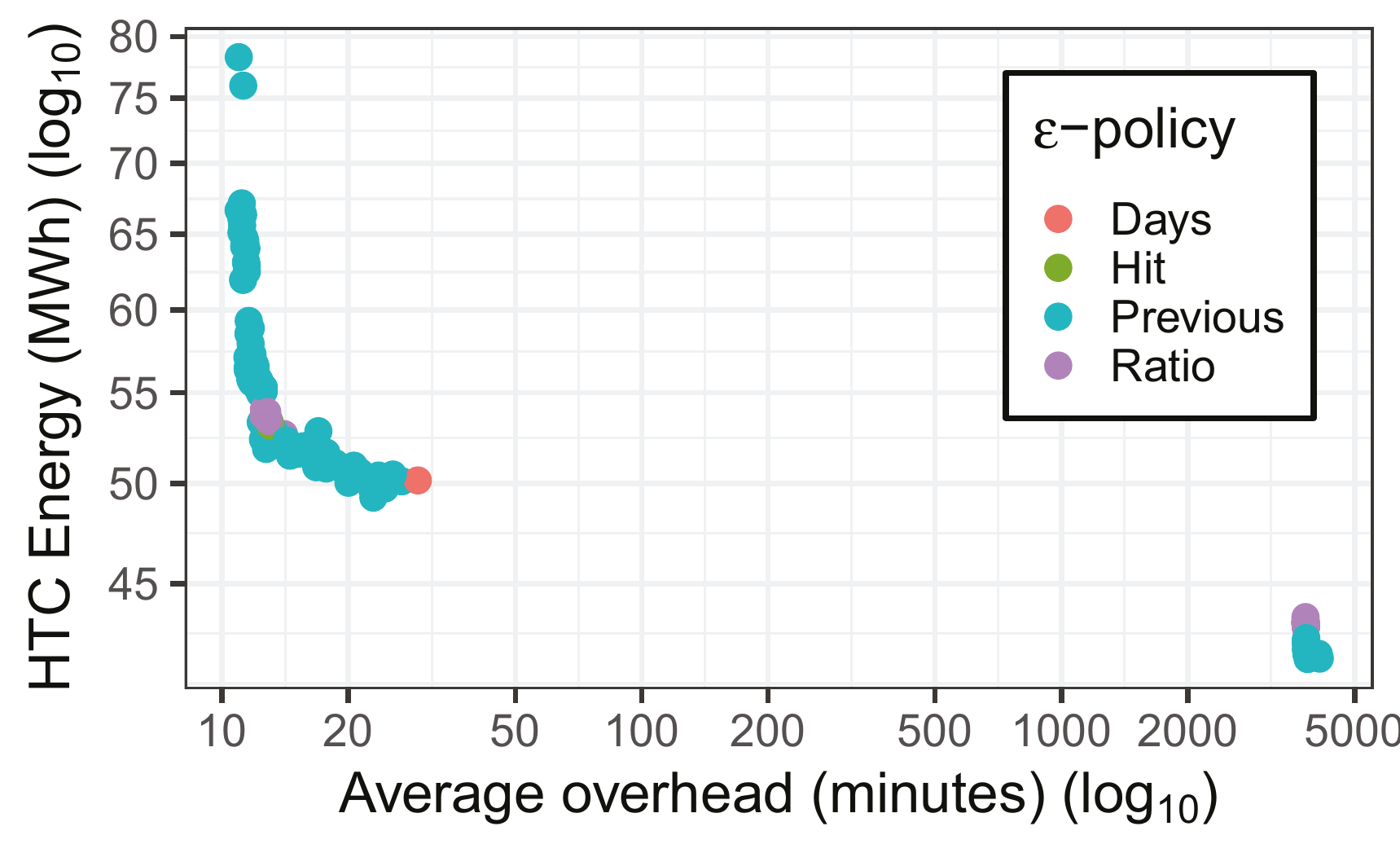}
\hfil
\includegraphics[width=0.3\linewidth]{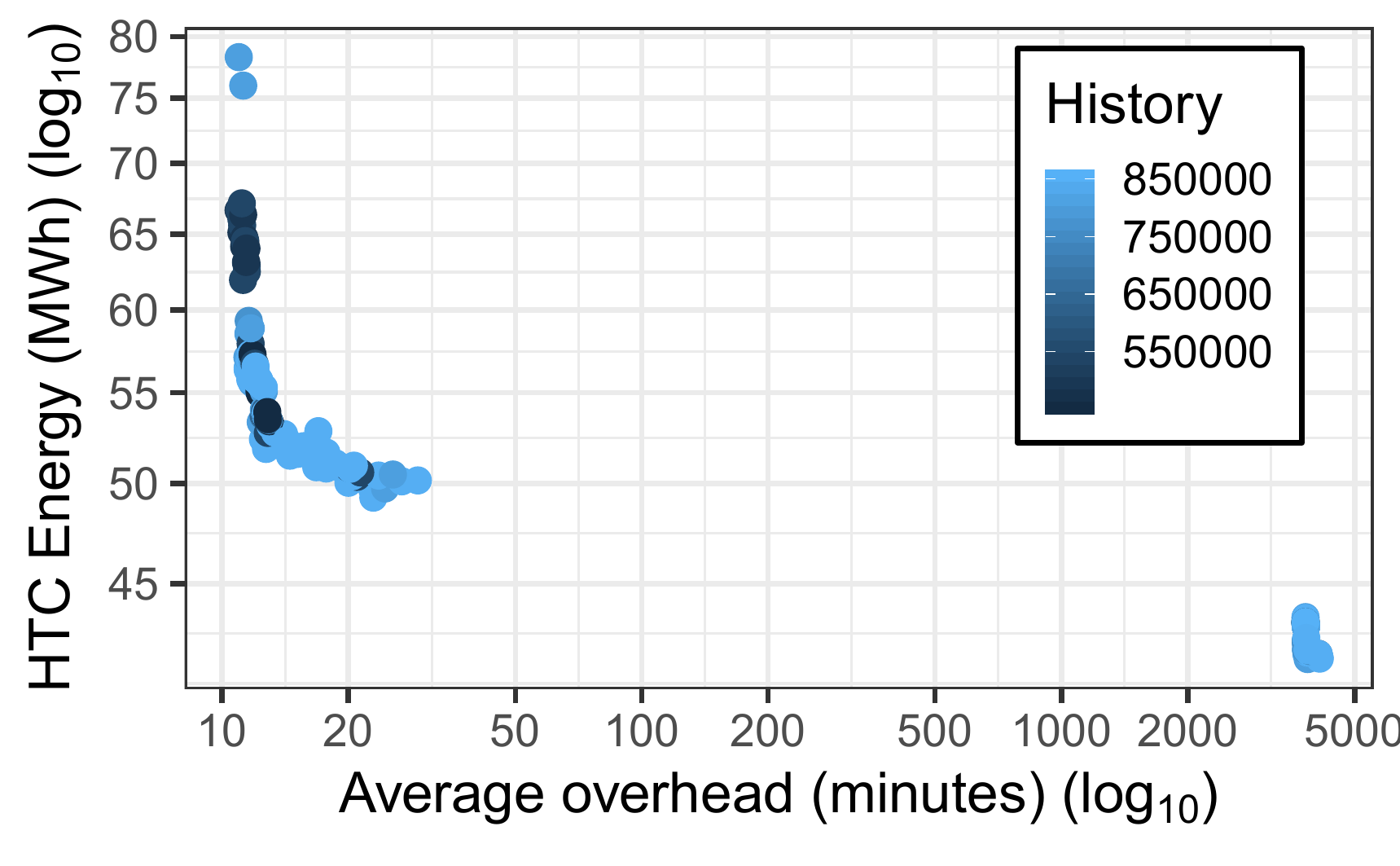}
\hfil
\includegraphics[width=0.3\linewidth]{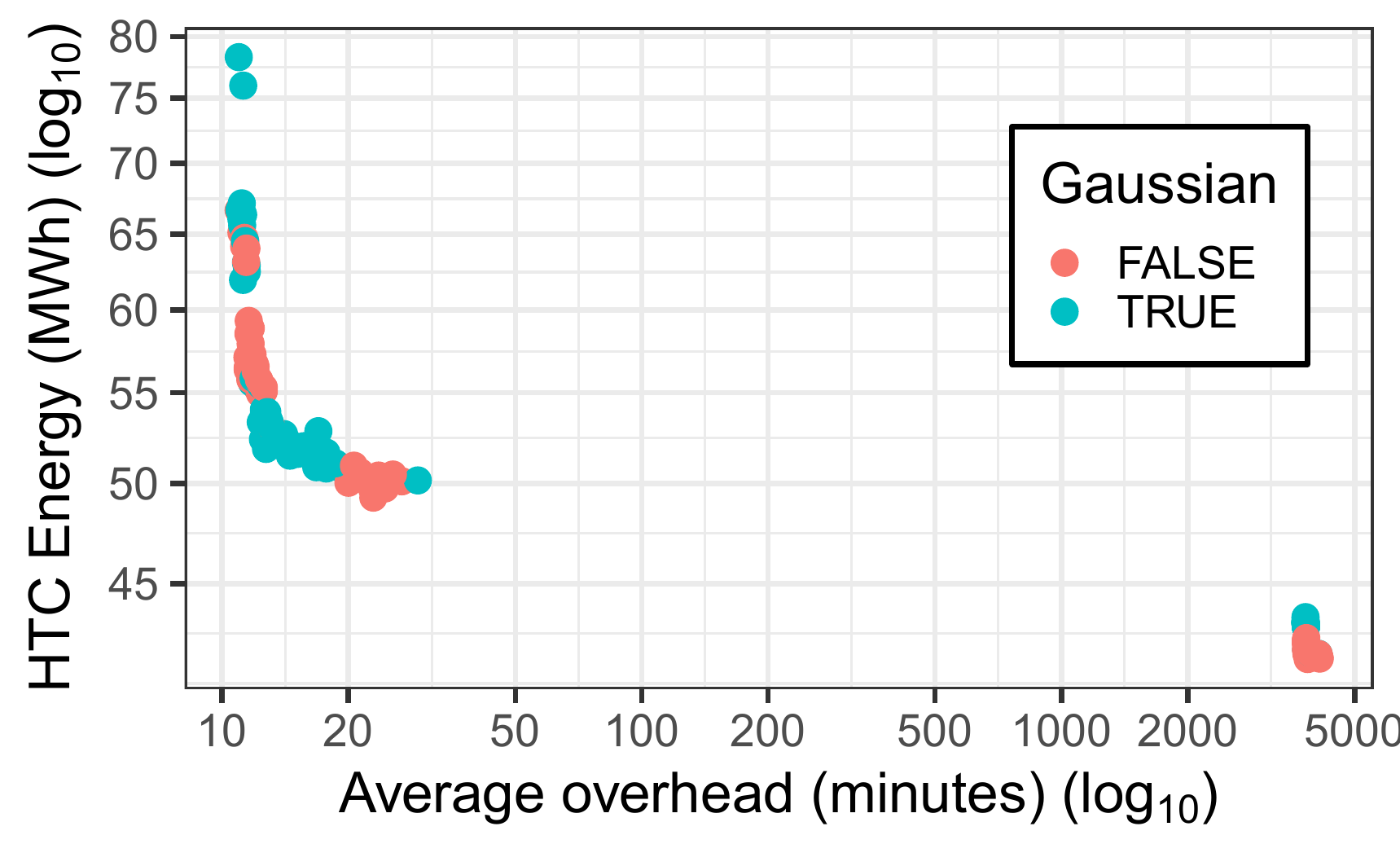}
  \caption{a) $\epsilon$-policy, b) Reward history window size and c) Using a gaussian decay over the reward history window}
  \label{fig:gaussian}
\end{figure*}
\begin{figure*}[!t]
\includegraphics[width=0.3\linewidth]{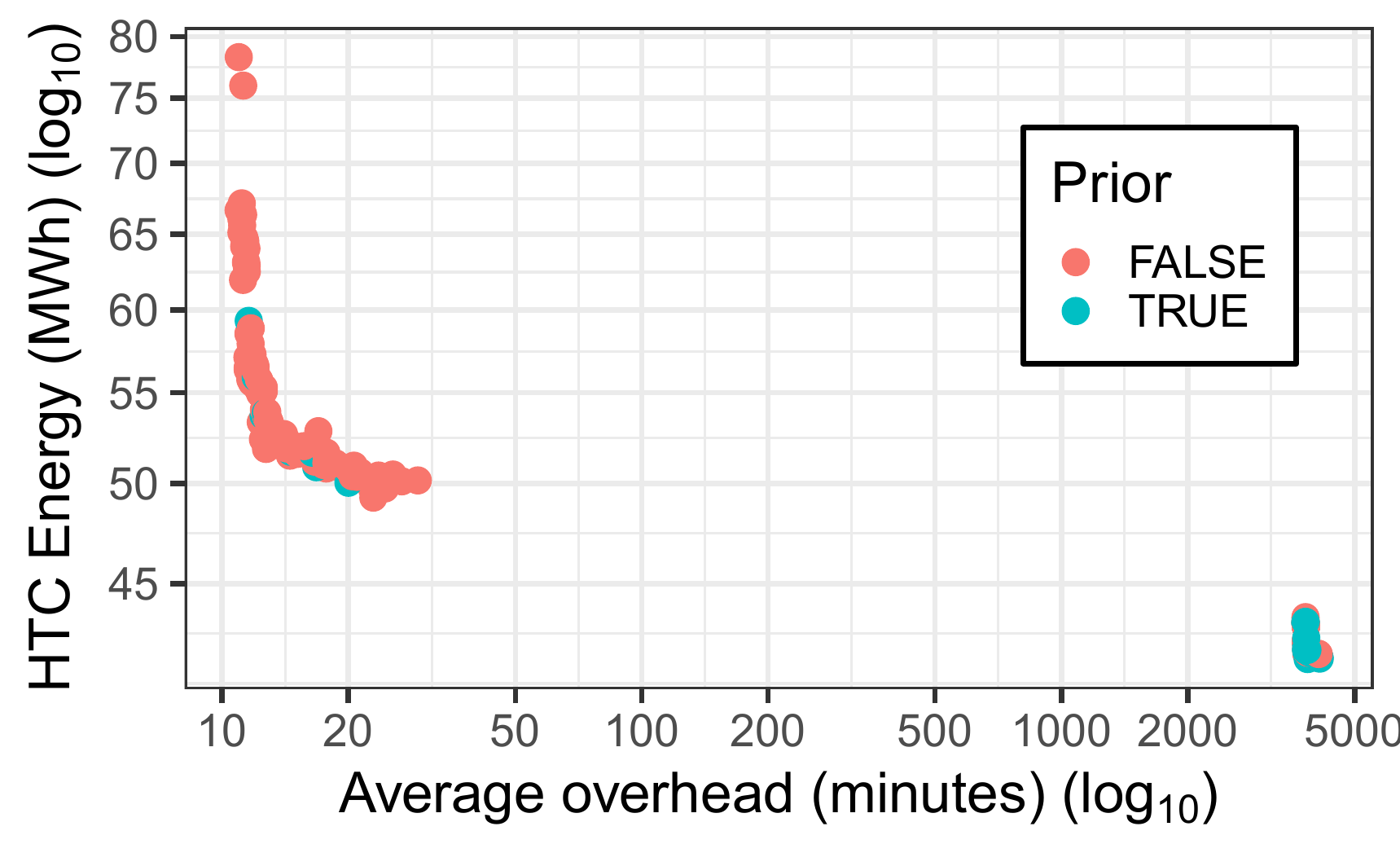}
\includegraphics[width=0.3\linewidth]{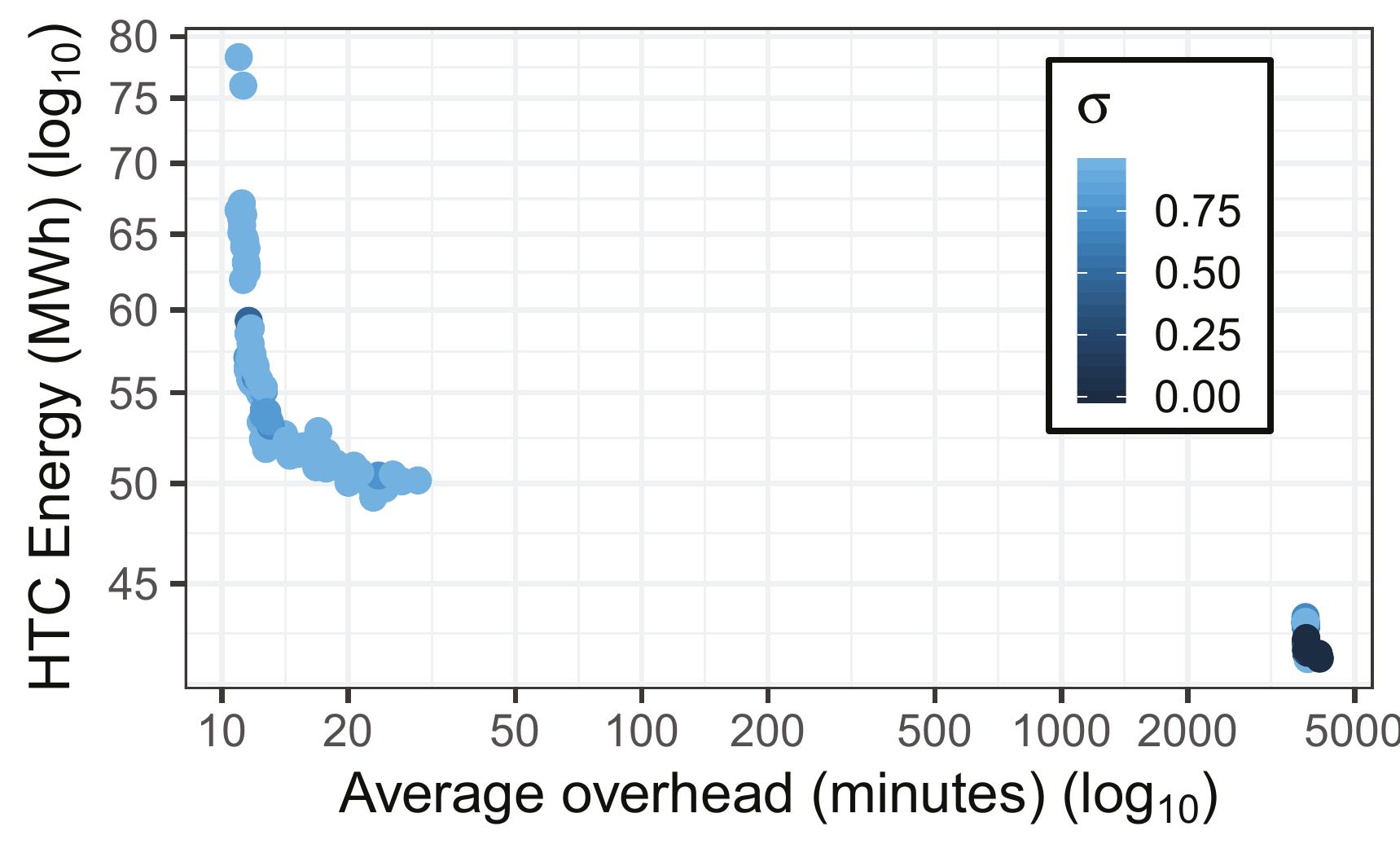}
\hfil
\includegraphics[width=0.3\linewidth]{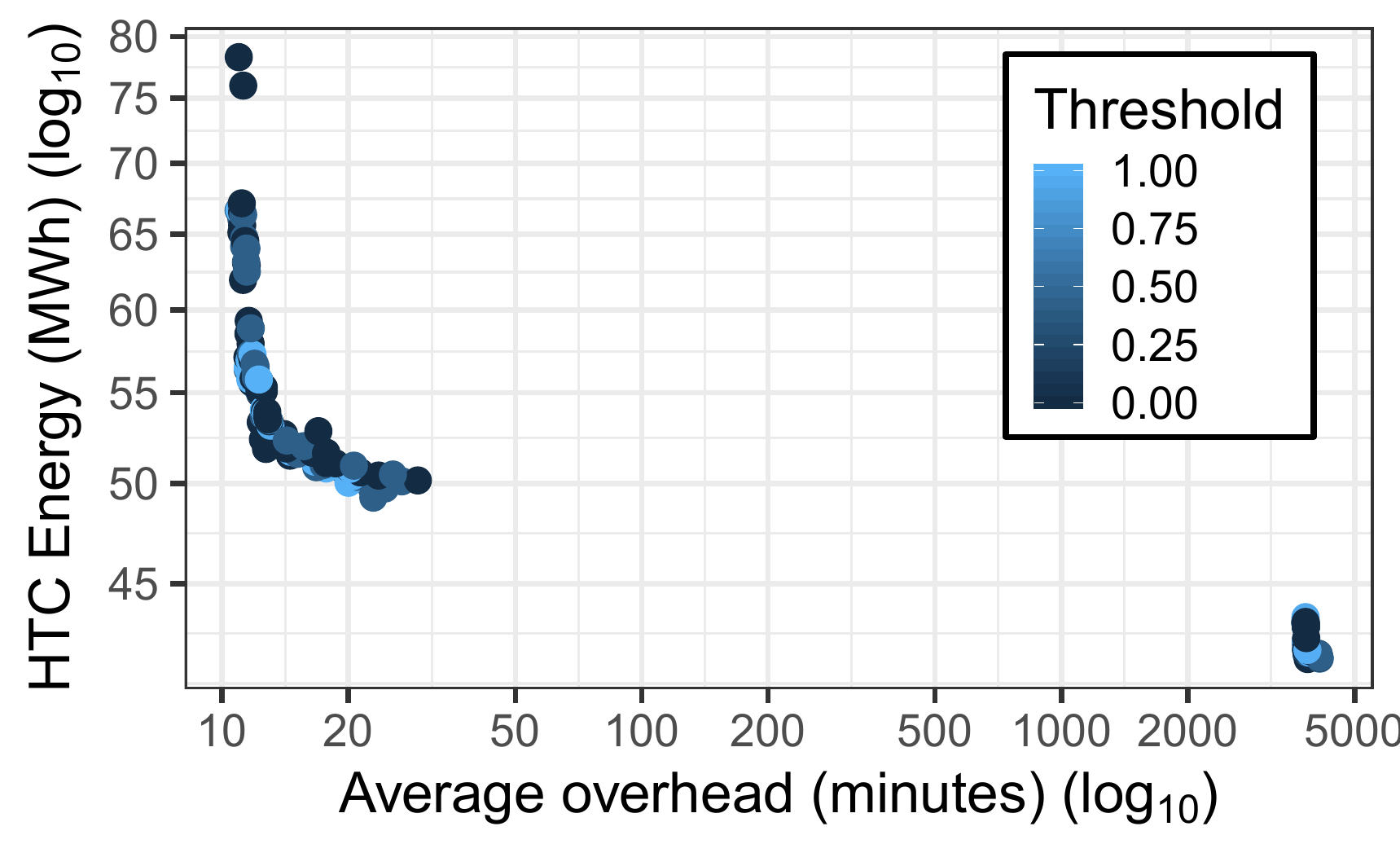}
  \caption{a) Impact of negative prior results, b) Impact of energy efficiency of computer and  c) Threshold impact}
  \label{fig:threshold}
\end{figure*}

Figure \ref{fig:changes} demonstrates the impact of adding $\delta$ to $\epsilon$ if the RL trainer has been running for less than the prescribed number of days. Here the GA has learnt to use large values of $\delta$ when energy consumption is more important, suggesting a more explorative approach favours energy efficiency -- perhaps due to the fact that over the simulation period the state varies significantly. By contrast, the number of days shows no clear pattern. Though as the $\delta$ value is often very small this may have little if any impact.

The granularity of the RL action space with respect to computers is presented in Figure \ref{fig:granularity}a. In almost all cases cluster level is the best choice. This is most likely a consequence of the fact that it is a compromise between fine-grained computer level and course-grained whole system level. Interestingly for minimum energy whole system becomes more optimal. Perhaps a consequence of most tasks being held in a queue rather than executed, hence more fine-grained knowledge no longer helps.

The other aspect of action/state space -- day or week -- is presented in Figure \ref{fig:granularity}b. Here all but the most extreme overhead cases are optimal with the day case. This would suggest that, although there is a difference based on the day of the week, this can only be exploited in the case where energy reduction is key.
%
%

The $\epsilon$ policy is compared in Figure \ref{fig:gaussian}a. In almost all cases the `Previous' policy is optimal apart from a small number of cases. This indicates that basing $\epsilon$ on the average reward of the previous day is the best policy. The ratio of best reward to average reward makes up most of the remaining points indicating that for both of these cases an adaptive policy which can move between explorative and exploitative modes over time is the best approach -- a consequence of the state of the system changing as time progresses. Only one `static' policy is seen as optimal - where the value of $\epsilon$ changes by the number of days the RL has been running.

Reward history and applying gaussian decay over the history is presented in Figures \ref{fig:gaussian}b and \ref{fig:gaussian}c. History size seems to be bimodal with the extremes of overhead and energy  having a value around 840,000 whilst most of the low overhead values are in the region of 500,000. This suggests that forgetting history more quickly favours lower overheads -- but at the expense of higher energy consumption. By contrast, the choice of when to use a gaussian decay is less obvious suggesting that other factors are at play. 

Increasing $\epsilon$ by 0.1 when prior rewards are negative is a mixed case for central points -- Figure \ref{fig:threshold}a -- though for lowest overhead and lowest energy the best approach appears to be disabled and enabled respectively, again suggesting that a more explorative approach favours lower energy. The impact of taking computer energy efficiency into account when computing the reward is presented in Figure \ref{fig:threshold}b. One would assume that taking energy efficiency into account would be important for low overall energy usage, however, the opposite seems to be the case. This would suggest that for extremely low energy cases whether the task is launched or not is most important. The ratio of best reward to average reward is presented in Figure \ref{fig:threshold}c. Results are variable, but lower thresholds tend to be better.
\section{Conclusions}
\label{conc}
In this paper we have demonstrated the potential of genetic algorithms, specifically NSGA-II, for efficient design space exploration of the operating policies of digital twin simulations. We apply the approach to parameterise the operating policies of a target system, a digital twin simulation of a high-throughput computing infrastructure. We evaluate the performance of the system with respect to energy consumption and performance. Through this approach we are able to reduce energy consumed by an HTC system by optimising the parameters of a RL scheduler by ${\sim}36\%$ with only negligible increase to the overheads. This allows us to more efficiently tune the parameter sets in situations where there are more parameter combinations than can feasibly be searched, and multiple objectives over which to optimise. 

In future work we plan to optimise over an increased number of objectives such as turnaround time for individual users and maximum waiting time per task. We also hope to implement our findings in a real HTCondor system.








\bibliographystyle{IEEEtran}
%
\bibliography{arXiv}

%
%
%

\end{document}